\documentclass[twocolumn]{aastex63}
\usepackage[utf8]{inputenc}
\usepackage{units}
\usepackage{hyperref}
\usepackage{blindtext}

\submitjournal{ApJ}

\begin{document}

\title{A search for photons with energies above $\unit[2{\times}10^{17}]{eV}$ using hybrid data from the low-energy extensions of the Pierre Auger Observatory}

\author{P.~Abreu}
\affiliation{Laborat\'orio de Instrumenta\c{c}\~ao e F\'\i{}sica Experimental de Part\'\i{}culas -- LIP and Instituto Superior T\'ecnico -- IST, Universidade de Lisboa -- UL, Lisboa, Portugal}

\author{M.~Aglietta}
\affiliation{Osservatorio Astrofisico di Torino (INAF), Torino, Italy}
\affiliation{INFN, Sezione di Torino, Torino, Italy}

\author{J.M.~Albury}
\affiliation{University of Adelaide, Adelaide, S.A., Australia}

\author{I.~Allekotte}
\affiliation{Centro At\'omico Bariloche and Instituto Balseiro (CNEA-UNCuyo-CONICET), San Carlos de Bariloche, Argentina}

\author{K.~Almeida Cheminant}
\affiliation{Institute of Nuclear Physics PAN, Krakow, Poland}

\author{A.~Almela}
\affiliation{Instituto de Tecnolog\'\i{}as en Detecci\'on y Astropart\'\i{}culas (CNEA, CONICET, UNSAM), Buenos Aires, Argentina}
\affiliation{Universidad Tecnol\'ogica Nacional -- Facultad Regional Buenos Aires, Buenos Aires, Argentina}

\author{J.~Alvarez-Mu\~niz}
\affiliation{Instituto Galego de F\'\i{}sica de Altas Enerx\'\i{}as (IGFAE), Universidade de Santiago de Compostela, Santiago de Compostela, Spain}

\author{R.~Alves Batista}
\affiliation{IMAPP, Radboud University Nijmegen, Nijmegen, The Netherlands}

\author{J.~Ammerman Yebra}
\affiliation{Instituto Galego de F\'\i{}sica de Altas Enerx\'\i{}as (IGFAE), Universidade de Santiago de Compostela, Santiago de Compostela, Spain}

\author{G.A.~Anastasi}
\affiliation{Osservatorio Astrofisico di Torino (INAF), Torino, Italy}
\affiliation{INFN, Sezione di Torino, Torino, Italy}

% ...

\collaboration{1000}{The Pierre Auger Collaboration}
\email{spokespersons@auger.org}

\begin{abstract}
Ultra-high-energy photons with energies exceeding $\unit[10^{17}]{eV}$ offer a wealth of connections to different aspects of cosmic-ray astrophysics as well as to gamma-ray and neutrino astronomy. The recent observations of photons with energies in the $\unit[10^{15}]{eV}$ range further motivate searches for even higher-energy photons. In this paper, we present a search for photons with energies exceeding $\unit[2{\times}10^{17}]{eV}$ using about 5.5 years of hybrid data from the low-energy extensions of the Pierre Auger Observatory. The upper limits on the integral photon flux derived here are the most stringent ones to date in the energy region between $10^{17}$ and $\unit[10^{18}]{eV}$.
\end{abstract}


\section{Introduction}
\label{sec:introduction}

The recent observations of photons with energies of a few $\unit[10^{14}]{eV}$ from decaying neutral pions, both from a direction coincident with a giant molecular cloud (HAWC J1825-134, \citet{Albert:2020yty}) and from the Galactic plane~\citep{Amenomori:2021gmk}, provide evidence for an acceleration of cosmic rays to energies of several $\unit[10^{15}]{eV}$, and above, in the Galaxy. A dozen of sources emitting photons with energies up to $\unit[10^{15}]{eV}$ have even been reported~\citep{LhaasoNature}, and in at least one of them~(LHAASO J2108+515, also in directional coincidence with a giant molecular cloud), these photons might have a hadronic origin~\citep{Cao:2021hdt}. Observations of these photons are key in probing the mechanisms of particle acceleration, completing the multi-messenger approach aimed at understanding the non-thermal processes producing cosmic rays. The detection of even higher-energy photons would be of considerable interest in discovering extreme accelerators in the Galaxy. Also, should one detect photons of such energies clustered preferentially in the direction of the Galactic Center, then this could highlight the presence of super-heavy dark matter produced in the early Universe and decaying today~(see, e.g., \citet{Berezinsky:1998rp,Benson:1999ie,Medina-Tanco:1999ktg,Aloisio:2006yi,Siffert:2007zza,Kalashev:2016cre,Alcantara:2019sco}).

Above $\unit[10^{17}]{eV}$, the absorption length for photons almost matches the scale of the Galaxy, and reaches that of the Local Group as the energy increases~\citep{Risse:2007sd}. The observation of point-like sources of photons would be compelling evidence for the presence of ultra-high-energy accelerators within such a local horizon. Diffuse fluxes of photons are also expected from farther away from the interactions of ultra-high-energy cosmic rays (UHECRs) with the background photon fields permeating the extragalactic space~(see, e.g.,~\citet{Gelmini:2005wu,Kampert:2011hkm,Bobrikova:2021kuj}) or with the interstellar matter in the Galactic disk~\citep{Berat2021}. Although the estimation of these cosmogenic photon fluxes suffers from several uncertainties of astrophysical origin, such as, in particular, the exact composition of UHECRs, they can be determined to range, at most, around $\unit[10^{-2}]{km^{-2}\,sr^{-1}\,yr^{-1}}$ above $\unit[10^{17}]{eV}$ and around $\unit[10^{-3.5}]{km^{-2}\,sr^{-1}\,yr^{-1}}$ above $\unit[10^{18}]{eV}$. These cosmogenic fluxes are more than two orders of magnitude below the sensitivity of current instruments, thereby constituting a negligible background for detecting photons from point sources, extended structures, or exotic phenomena.

Previous searches for a diffuse flux of photons using data from KASCADE-Grande~\citep{Apel:2017ocm} and EAS-MSU~\citep{Fomin:2017ypo} have led to upper limits on photon fluxes of the order of $\unit[10]{km^{-2}\,sr^{-1}\,yr^{-1}}$ for energy thresholds between $\unit[10^{17}]{eV}$ and $\unit[3{\times}10^{17}]{eV}$, while at higher energies, at a threshold of $\unit[10^{18}]{eV}$, upper limits of the order of $\unit[10^{-2}]{km^{-2}\,sr^{-1}\,yr^{-1}}$ were determined using data from the Pierre Auger Observatory~\citep{Savina:2021zrn}. The aim of the study reported in this paper is to search for primary photons with energies above $\unit[2{\times}10^{17}]{eV}$ using data from the low-energy extensions of the Pierre Auger Observatory, which are briefly presented in Section~\ref{sec:auger}. The data set used in this study is described in Section~\ref{sec:samples} together with the simulations needed to establish the selection criteria aimed at distinguishing photon-induced air showers from those initiated by hadronic cosmic rays. In Section~\ref{sec:analysis}, the specificities of the photon-induced showers are used to define discriminating observables, which are then combined to search for photon candidate events in the data. Results are given in Section~\ref{sec:results} and, from the absence of a photon signal, upper limits on the integral photon flux are derived that improve the previous ones mentioned before. Finally, the astrophysical significance of these limits is discussed in Section~\ref{sec:discussion}.

\section{The Pierre Auger Observatory}
\label{sec:auger}

\begin{figure}[t]
\plotone{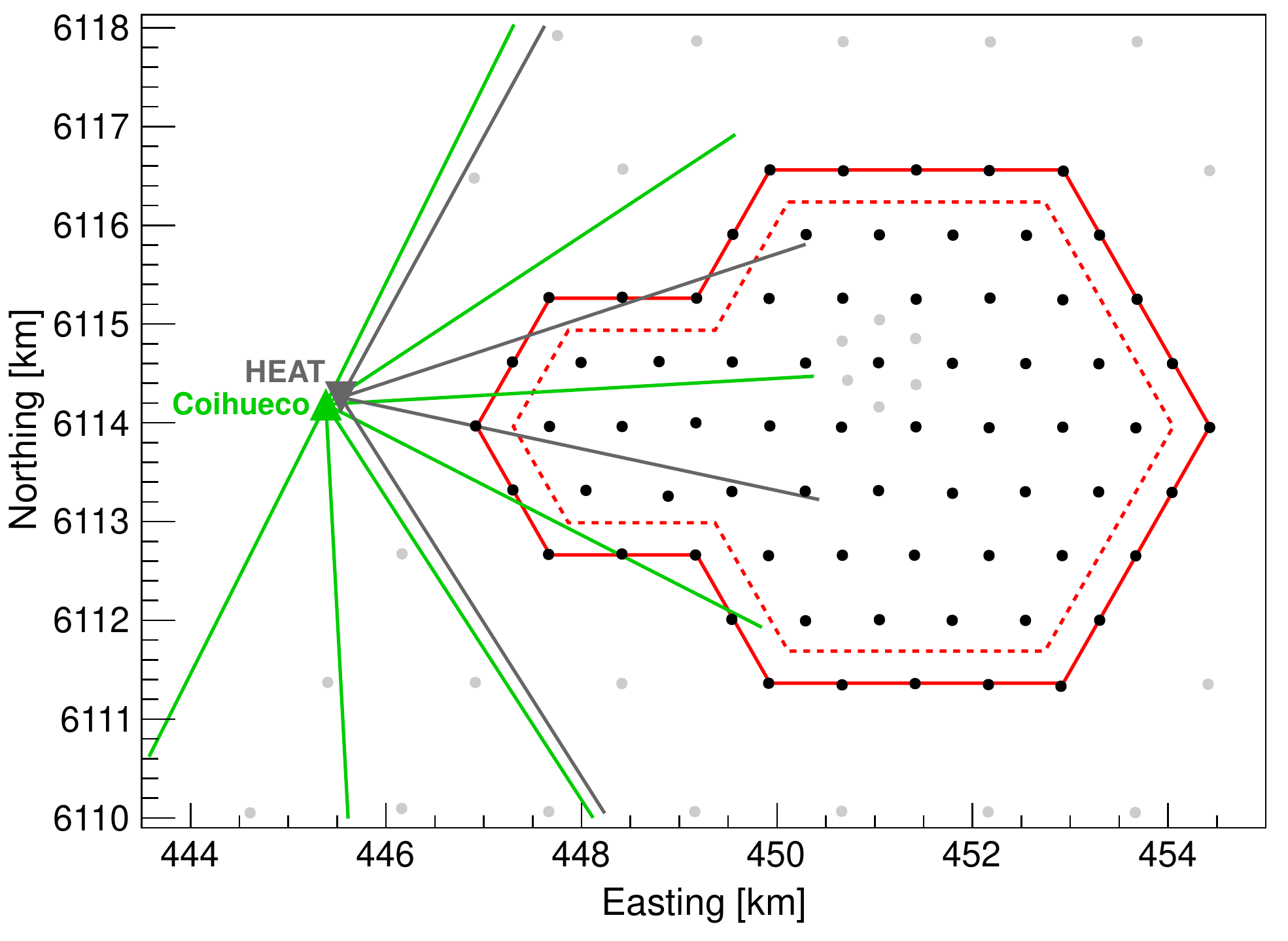}
\caption{
Schematic depiction of the part of the detector layout of the Pierre Auger Observatory~\citep{ThePierreAuger:2015rma} that is relevant for the analysis discussed here. Detector stations from the $\unit[750]{m}$ SD array are shown as black points. Detector stations that are not used in this analysis (for example from the $\unit[1500]{m}$ SD array) are greyed out. The projections of the fields-of-view of the fluorescence telescopes from Coihueco and HEAT  on the ground are indicated by the green and gray lines, respectively. Note that Coihueco and HEAT cover different elevation ranges. The outline of the $\unit[750]{m}$ SD array is given by the solid red line, while the dashed red line marks the region where the shower core of an air shower event has to be located for the event to be accepted in this analysis (see Sec.~\ref{sec:samples}).
}
\label{fig:detectorlayout}
\end{figure}

The Pierre Auger Observatory~\citep{ThePierreAuger:2015rma}, located near Malarg\"ue, Argentina, offers an unprecedented exposure for UHE photons. A key feature of the Pierre Auger Observatory is the hybrid concept, combining a Surface Detector array (SD) with a Fluorescence Detector (FD). The SD consists of 1600 water-Cherenkov detectors arranged on a triangular grid with a spacing of $\unit[1500]{m}$, covering a total area of $\unit[3000]{km^2}$. The SD is overlooked by 24 fluorescence telescopes, located at four sites at the border of the array. The SD samples the lateral shower profile at ground level, i.e., the distribution of particles as a function of the distance from the shower axis, with a duty cycle of $\unit[{\sim}100]{\%}$, while the FD records the longitudinal shower development in the atmosphere above the SD. The FD can only be operated in clear, moonless nights, reducing the duty cycle to $\unit[{\sim}15]{\%}$. Through combining measurements from both detector systems in hybrid events, a superior accuracy of the air-shower reconstruction can be achieved than with just one system. In the western part of the SD array, 50 additional SD stations have been placed between the existing SD stations, forming a sub-array with a spacing of $\unit[750]{m}$ and covering a total area of about $\unit[27.5]{km^2}$. With this sub-array, air showers of lower primary energy (below $\unit[10^{18}]{eV}$) with a smaller footprint can be measured. To allow also for hybrid measurements in this energy range, where air showers develop above the field of view of the standard FD telescopes, three additional High-Elevation Auger Telescopes (HEAT) have been installed at the FD site Coihueco, overlooking the $\unit[750]{m}$ SD array. The HEAT telescopes operate in the range of elevation angles from $30^\circ$ to $60^\circ$, complementing the Coihueco telescopes operating in the $0^\circ$ to $30^\circ$ range. The combination of the data from both HEAT and Coihueco (``HeCo'' data) enable fluorescence measurements of air showers over a large range of elevation angles. A schematic depiction of the detector layout, including the $\unit[750]{m}$ array and HEAT, can be found in Fig.~\ref{fig:detectorlayout}.

\section{Data Samples and Simulations}
\label{sec:samples}

The analysis is based on hybrid data collected by the Coihueco and HEAT telescopes and the $\unit[750]{m}$ SD array between 1 June 2010 and 31 December 2015. Subsequent data will be used in a follow-up paper. In the present paper, we use the same analysis techniques as in~\citet{Aab:2016agp} to provide a first search for photons in the energy range between $\unit[2{\times}10^{17}]{eV}$ and $\unit[10^{18}]{eV}$ using data from the Pierre Auger Observatory. The follow-up paper will not only make use of a larger dataset, but also profit from an analysis that is tailor-made for the low-energy enhancements of the Observatory.

Several selection criteria are applied to this dataset to ensure a good reconstruction of the air-shower events and a reliable measurement of the observables used to discriminate photon- and hadron-induced air showers (see Sec.~\ref{sec:analysis}). These criteria are summarized in the following.

The total dataset contains $587{,}475$ ``HeCo'' events at the detector level, before any further selection criteria are applied. A sub-sample consisting of about $\unit[5]{\%}$ of the total dataset ($29{,}531$ events, selected from the full data period using the simple prescription $T_{\text{GPS}}\,\text{mod}\,20=0$, where $T_{\text{GPS}}$ denotes the time the event was recorded in units of GPS seconds) was used as a ``burnt sample'' to optimize the event selection and perform cross-checks on the analysis. The events from the ``burnt sample'' are not used in the final analysis.

At the geometry level of the event selection, it is required that the events are reconstructed using the hybrid event reconstruction procedure, taking into account the timing information from a triggered SD station from the $\unit[750]{m}$ SD array in addition to the FD measurements. To exclude events pointing directly towards the FD telescope, where Cherenkov light will distort the FD measurement, a minimum viewing angle of $15^\circ$ is required. Lastly, only events where the shower core is reconstructed within the inner region of the $\unit[750]{m}$ SD array (marked by the dashed red line in Fig.~\ref{fig:detectorlayout}) and where the zenith angle is below $60^\circ$ are considered. More inclined events are not taken into account because of the absorption of the electromagnetic component of the air showers in the atmosphere and the resulting smaller trigger efficiency at lower energies.

At the third level of the event selection, the profile level, events with an unreliable reconstruction of the longitudinal profile of the air shower are discarded using a cut based on the reduced $\chi^2$ of the fit of a Gaisser-Hillas function to the recorded profile. Events are only accepted when the reconstructed atmospheric depth of the shower maximum $X_{\text{max}}$ is inside the geometrical field of view of the fluorescence telescopes and gaps in the recorded tracks, which can appear, for example, for air showers crossing several telescopes, amount to less than $\unit[30]{\%}$ of the total observed track length. Finally, it is required that the uncertainty on the reconstructed photon energy $E_\gamma$, defined as the calorimetric energy taken from the integration of the profile plus a missing-energy correction of $\unit[1]{\%}$ appropriate for primary photons~\citep{Aab:2016agp}, is less than $\unit[20]{\%}$.

Since the precise knowledge of the atmospheric conditions is crucial for the hybrid reconstruction, events recorded during periods without information on the aerosol content of the atmosphere are not taken into account. To exclude events where the recorded profile may be distorted due to clouds over the Observatory, only events from known cloud-free periods are accepted. Events where no information on the cloud coverage is available from either the Lidar system installed at the FD site Coihueco~\citep{BenZvi:2006xb} or infrared data from the GOES-12 satellite~\citep{PierreAuger:2013lgb} are excluded.

Finally, the last selection criterion removes events where fewer than four of the six SD stations in the first $\unit[750]{m}$ hexagon around the station with the largest signal are active. Such cases can occur, e.g., in the border region of the array or when individual SD stations are temporarily offline and not taking data. In this case, the discriminating observables $S_b$ and $N_{\text{stations}}$ (see Sec.~\ref{sec:analysis}) can be underestimated, mimicking air showers initiated by photons.

The numbers of events after each level of the event selection and the associated selection efficiencies are given in Tab.~\ref{tab:eventselection}, excluding the ``burnt sample'' as mentioned before. The largest reduction occurs already at the geometry level. Here, the main contribution comes from the restriction of the acceptance to the area of the $\unit[750]{m}$ SD array, followed by the requirement that the events have to be reconstructed using the hybrid procedure. 
After all cuts, $2{,}204$ events remain with a photon energy $E_\gamma$ above $\unit[2{\times}10^{17}]{eV}$.

\begin{table}[t]
\centering
\caption{
Numbers of events from the data sample (excluding the ``burnt sample'') passing the different event selection levels and the associated selection efficiencies relative to the preceding level. See the text for explanations.
}
\begin{tabular}{rrr}
\hline\hline
Total number of ``HeCo'' events: & $557{,}944$ & ---\phantom{\,\%} \\
After geometry level: & $20{,}545$  & \unit[3.7]{\%}\\
After profile level: & $12{,}129$ & \unit[59.0]{\%}\\
After atmosphere level: & $4{,}373$ & \unit[36.1]{\%}\\
After $S_b$ level: & $3{,}873$ & \unit[88.6]{\%}\\
$E_\gamma \geq
\unit[2{\times}10^{17}]{eV}$: & $2{,}204$ & \unit[56.9]{\%}\\
\hline\hline       
\end{tabular}
\label{tab:eventselection}
\end{table}

A large sample of simulated events has been used to study the photon/hadron separation by the observables used in this analysis, to train the multivariate analysis, and to evaluate its performance. Air-shower simulations have been performed with CORSIKA~\citep{Heck:1998vt}, using EPOS LHC~\citep{Pierog:2013ria} as the hadronic interaction model. About $72{,}000$ photon-induced and $42{,}000$ proton-induced air showers in six bins of equal width in $\log_{10}(E\,\unit{[eV]})$ between $\unit[10^{16.5}]{eV}$ and
$\unit[10^{19.5}]{eV}$, following a power-law spectrum with spectral index $-1$ within each bin, have been used. Zenith and azimuth angles of the simulated events were drawn from an isotropic distribution between $0$ and $65^\circ$ and from a uniform distribution between $0$ and $360^\circ$, respectively. Although they do not have a significant impact on the development of photon-induced air showers at the target energy range below $\unit[10^{18}]{eV}$, pre-showering~\citep{Erber:1966vv, Mcbreen:1981yc,Homola:2006wf} and LPM effects~\citep{Landau:1953um,Migdal:1956tc} were included in the simulations. Only proton-induced air showers are used as background, as these are the most ``photon-like'' compared to air showers induced by heavier nuclei such as helium.  Even though there are indications that the composition of UHECRs is getting heavier with energy (see, e.g.,~\citet{yushkov:2019taz}), the assumption of a pure-proton background in the context of a search for UHE photons can be taken as a conservative ``worst-case'' assumption, since including heavier nuclei  would always lead to a smaller estimate for the  contamination in the final sample of photon candidate events.

All simulated air-shower events are processed with the Auger Offline Software Framework~\citep{Argiro:2007qg} for a detailed simulation of the detector response. In these simulations, the actual detector status of both the SD and the FD as well as the atmospheric conditions at any given time during the aforementioned data period are taken into account, leading to a realistic estimate of the detector response. Each simulated air shower is used five times, each time with a different impact point on the ground, randomly taken from a uniform distribution encompassing the region of the $\unit[750]{m}$ SD array, and with a different event time, which was randomly determined according to the on-time of the Coihueco and HEAT telescopes during the data period used in this analysis. All simulated events are finally passed through the same event selection as the events from the data sample. After the event selection stage, the simulated samples contain about $55{,}000$ photon-induced events and about $35{,}000$ proton-induced events.

\section{Analysis}
\label{sec:analysis}

\begin{figure*}[t]
\gridline{
\fig{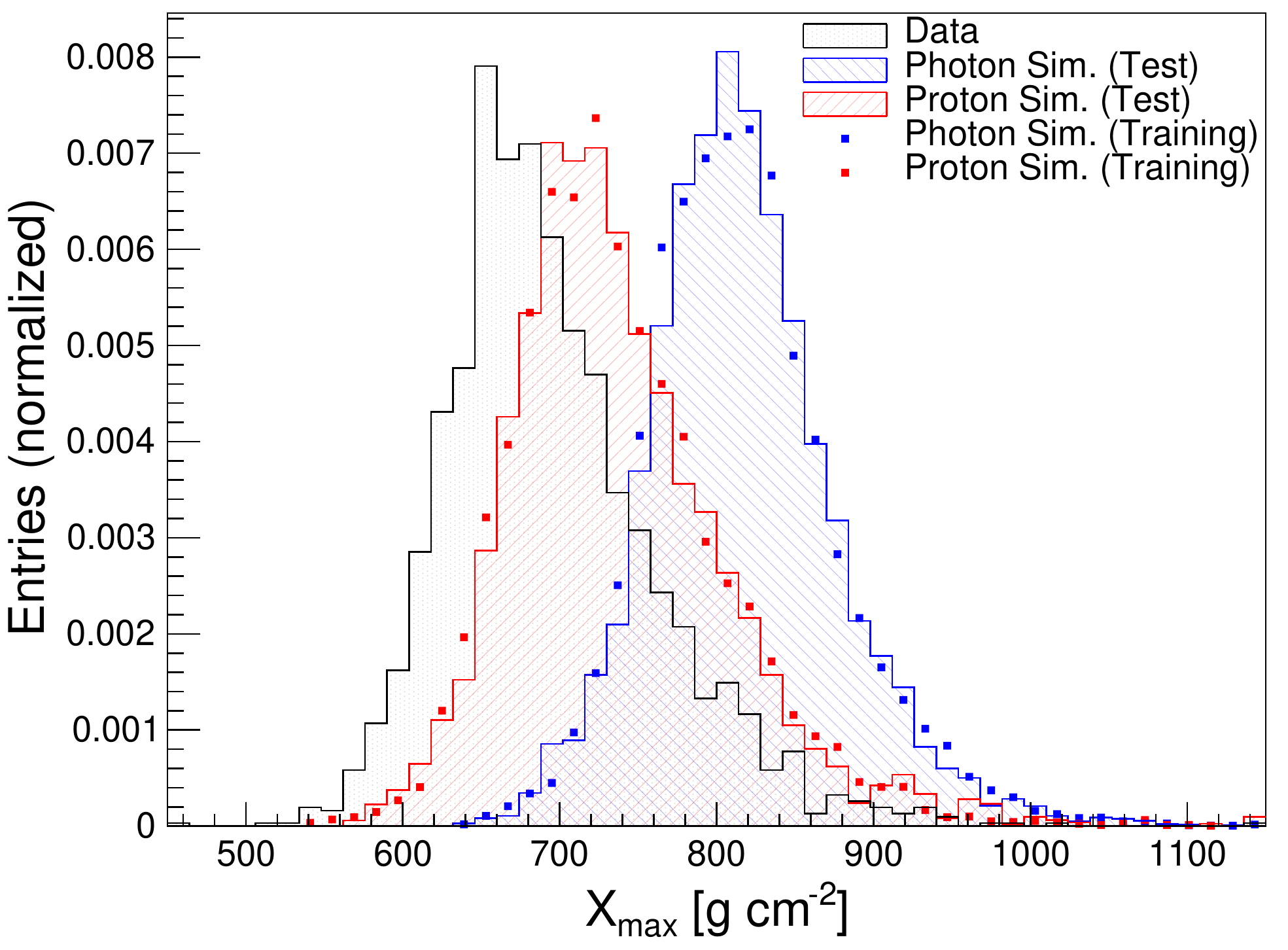}{0.32\textwidth}{(a)}
\fig{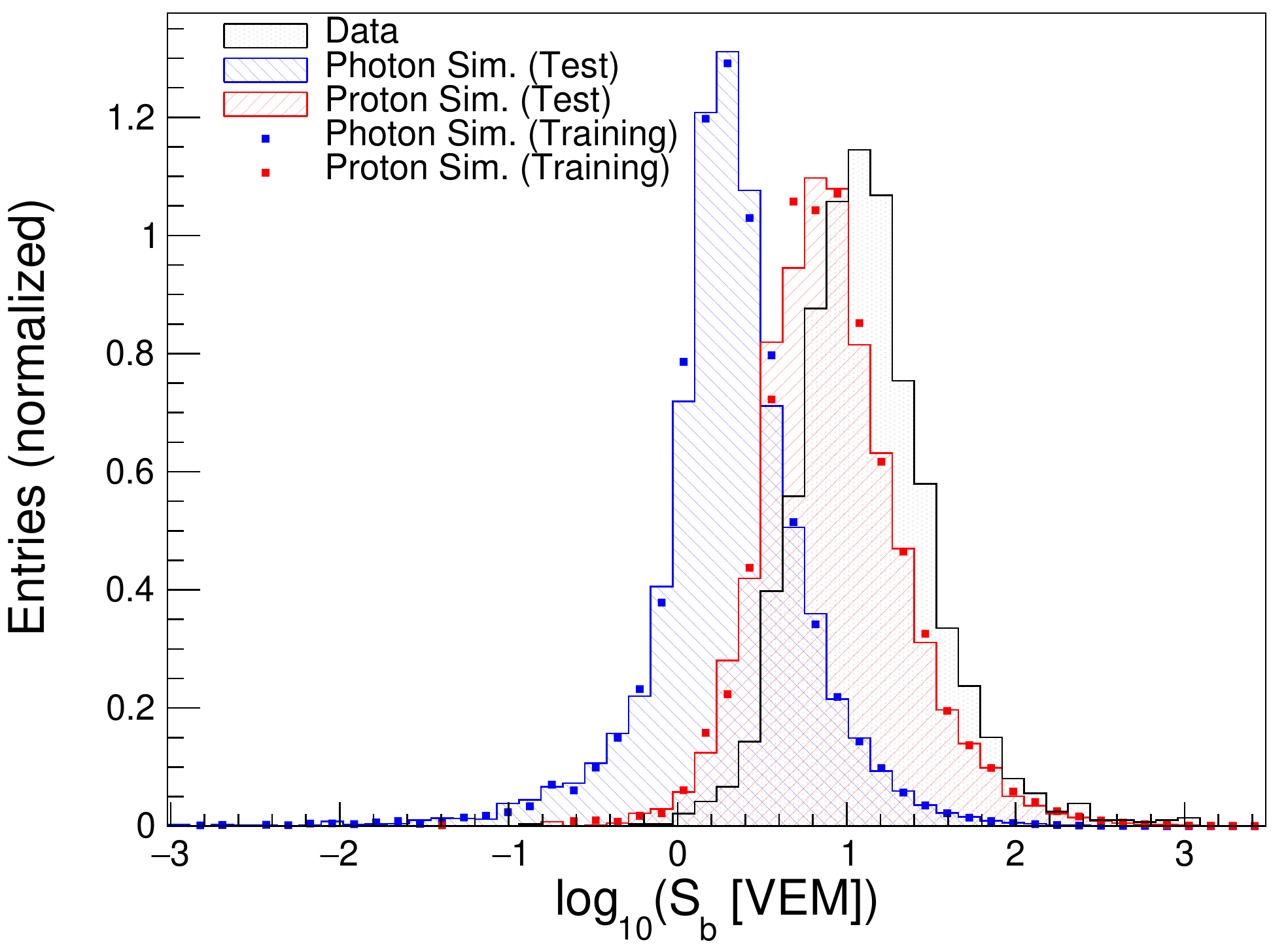}{0.32\textwidth}{(b)}
\fig{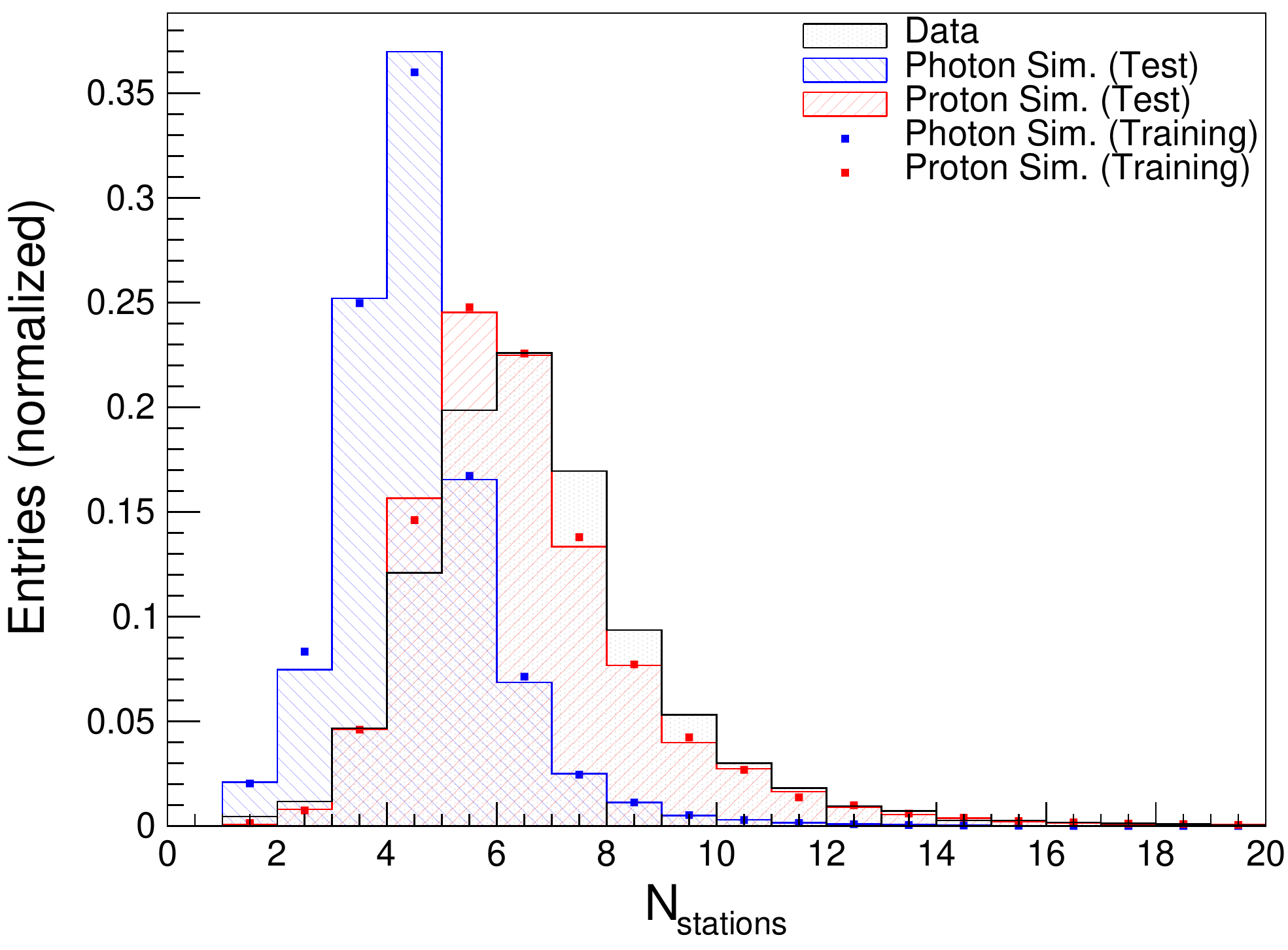}{0.32\textwidth}{(c)}
}
\caption{
Normalized distributions of the three discriminating observables $X_{\text{max}}$, $S_b$ and $N_{\text{stations}}$. The photon sample is shown in blue, the proton sample in red, and the data sample in black. Only events with $E_\gamma{>}\unit[2{\times}10^{17}]{eV}$ are shown. The simulated samples are subdivided into a training sample used to train the MVA and a test sample used to determine the separation power of the individual observables. Note that for illustrative purposes and to facilitate the comparison of the data distributions to the ones obtained from the simulated samples, the latter were weighted with an $E_{\gamma}^{-3}$ spectrum instead of the $E_{\gamma}^{-2}$ one used in the MVA (see Sec.~\ref{sec:analysis}).
}
\label{fig:distributions}
\end{figure*}

The search for primary photons presented in this work exploits the well-known differences in air-shower development for photon-induced and hadron-induced air showers: on the one hand, air showers initiated by photons develop deeper in the atmosphere than those initiated by hadrons, and on the other hand, they exhibit a smaller number of muons at ground level~\citep{Risse:2007sd}. The first difference can be quantified through $X_{\text{max}}$, which can be directly measured with the FD. To complement the FD-observable $X_{\text{max}}$, we use another quantity determined from the data of the $\unit[750]{m}$ SD array, called $S_b$, which is defined as follows~\citep{Ros:2011zg}:
\begin{equation}
    S_b = \sum_i S_i \times \left(\frac{R_i}{\unit[1000]{m}}\right)^{b},
\end{equation}
where $S_i$ denotes the measured signal in the $i$-th SD station at a perpendicular distance $R_i$ to the shower axis. The parameter $S_b$ has been chosen here as $b=4$ to optimize the photon-hadron separation in accordance with~\citet{Aab:2016agp}. By construction, $S_b$ is sensitive to the lateral distribution, which in turn depends on the depth of the air-shower development in the atmosphere and the number of muons. Hence, $S_b$ can be used to distinguish photon- and hadron-induced air showers. In addition to $X_{\text{max}}$ and $S_b$, the number of triggered SD stations $N_{\text{stations}}$ is also used in the analysis, as it has been shown in~\citet{Aab:2016agp} that it can significantly improve the overall performance of the analysis. The distributions of $X_{\text{max}}$, $S_b$ and $N_{\text{stations}}$ are shown in Fig.~\ref{fig:distributions} for the simulated samples as well as the data sample.

To combine the three discriminating observables, a multivariate analysis (MVA) is performed using the Boosted Decision Tree (BDT) method as implemented by the TMVA package~\citep{Hocker:2007ht}. To take into account energy and zenith angle dependencies, the photon energy $E_\gamma$ and the zenith angle $\theta$ are also included in the MVA. The MVA is trained using two thirds of the simulated samples described before, while the remaining third is used to test the trained MVA for consistency and calculate the performance of the MVA with regard to photon/hadron separation. In Fig.~\ref{fig:distributions}, the training and test sub-samples are denoted by the markers and the shaded regions, respectively, for both the photon and the proton samples. In the training and testing stages of the MVA, events are weighted according to a power-law spectrum $E^{-\Gamma}$ with a spectral index $\Gamma=2$, as in previous photon searches (see, e.g.,~\citet{Aab:2016agp}).

The distribution of the output from the BDT $\beta$, which is used as the final discriminator for separating photon-induced air showers from the hadronic background, is shown in Fig.~\ref{fig:bdt} for both the simulated and the data samples (see also Sec.~\ref{sec:results}). The photon and proton distributions are clearly separated. The background rejection at a signal efficiency of $\unit[50]{\%}$, i.e., the fraction of proton-induced events that have a $\beta$ larger than the median of the photon (test sample) distribution---which is used as the photon candidate cut, marked with the dashed line in Fig.~\ref{fig:bdt})---is $\unit[(99.87\,{\pm}\,0.03)]{\%}$, where the uncertainty has been determined through a bootstrapping method. When only events with $E_\gamma \geq \unit[2{\times}10^{17}]{eV}$ are taken into account,  the background rejection at $\unit[50]{\%}$ signal efficiency becomes $\unit[(99.91\,{\pm}\,0.03)]{\%}$, hence we expect a background contamination of $\unit[(0.09\,{\pm}\,0.03)]{\%}$. For the size of the data sample given in Tab.~\ref{tab:eventselection} ($2{,}204$ events), this would translate, under the assumption of a pure-proton background, to $1.98\,{\pm}\,0.66$ background events that are wrongly identified as photon candidate events. All of these numbers have been determined from the test samples (see above). Were the analysis to be based on $X_{\text{max}}$ only, the background rejection at $\unit[50]{\%}$ signal efficiency would be $\unit[92.5]{\%}$. The expected background contamination can therefore be reduced significantly by including the SD-related observables $S_b$ and $N_{\text{stations}}$.

\section{Results}
\label{sec:results}

Finally, we apply the analysis to the data sample to search for the presence of photon candidate events. The distributions of the three discriminating observables
$X_{\text{max}}$, $S_b$ and $N_{\text{stations}}$ for the data sample are shown in
Fig.~\ref{fig:distributions} together with the corresponding distributions for the simulated samples. In the following paragraphs, we briefly discuss these distributions.

The $X_{\text{max}}$ distribution for the data sample is shifted towards smaller $X_{\text{max}}$
values compared to the proton distribution. This is in line with current Auger results on the composition of ultra-high-energy cosmic rays: for example, in \citet{yushkov:2019taz}, the $\left<X_{\text{max}}\right>$ values that were measured above $\unit[10^{17.2}]{eV}$
are consistently below the expectation for primary protons, indicating a
heavier composition. As the average $X_{\text{max}}$ is
decreasing with increasing primary mass, a shift of the $X_{\text{max}}$
distribution for the data sample towards smaller
values is expected. Similarly, a composition effect can be seen in the $S_b$ and
$N_{\text{stations}}$ distributions. As the lateral shower profile gets
wider with increasing primary mass and the number of muons at ground
level increases, more
triggered SD stations are expected, on average,
compared to primary protons (and, consequently, primary
photons), leading to higher values of $N_{\text{stations}}$, as well as a higher
signal in these stations, which together with the higher multiplicity
leads to larger $S_b$ values. Also the choice of
the hadronic interaction model---here EPOS LHC---has an impact on the distributions
obtained for the simulated samples, in particular the proton
distributions. Furthermore,~\citet{PierreAuger:2014ucz,PierreAuger:2016nfk} indicate a possible
underestimation of the number of muons in simulations, which can also influence the distributions.

\begin{figure}[t]
\plotone{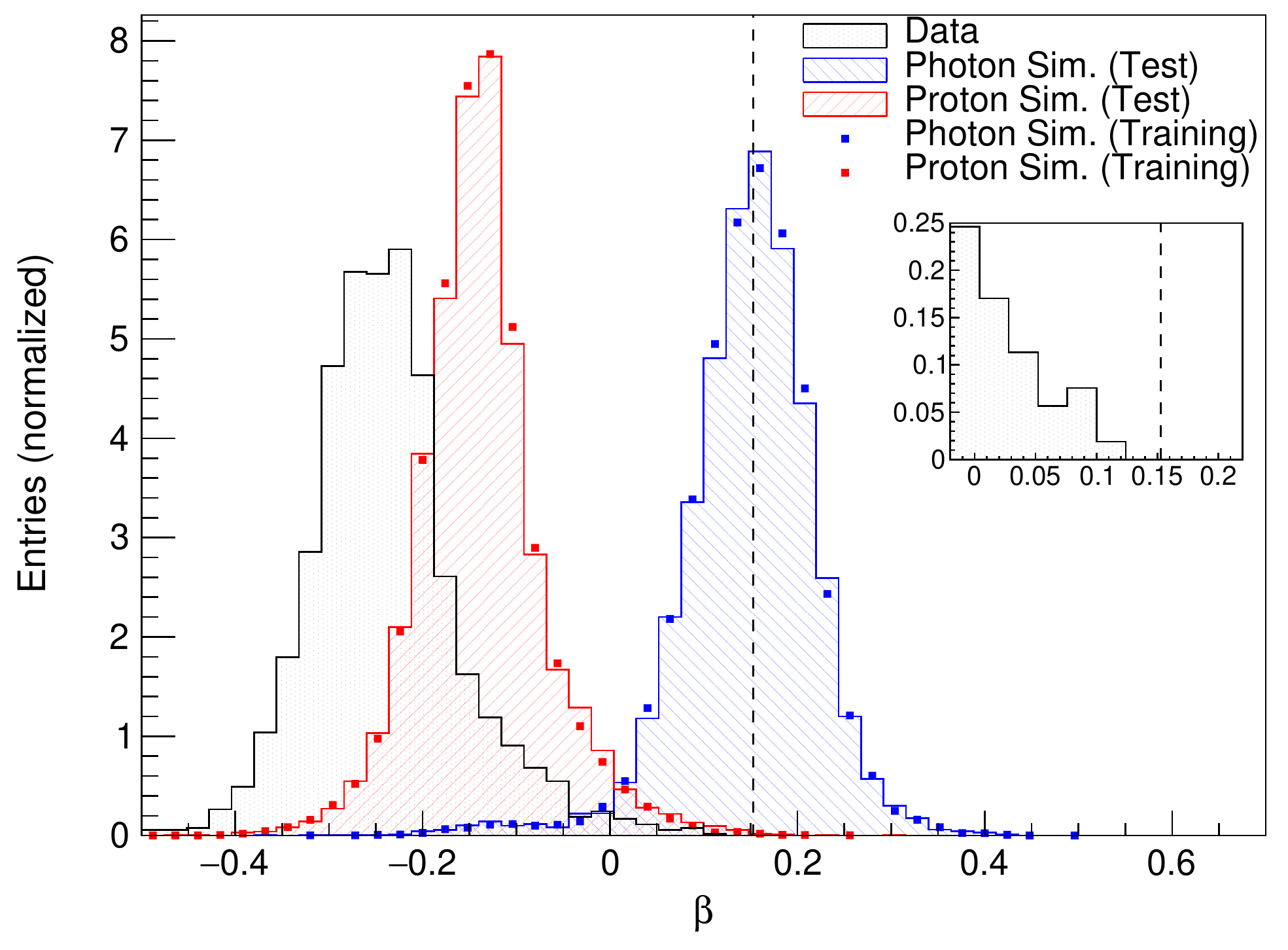}
\caption{
Normalized distributions of the final discriminator $\beta$. The photon sample is shown in blue, the proton sample in red, and the data sample in black. Only events with $E_\gamma{>}\unit[2{\times}10^{17}]{eV}$ are shown. The simulated samples are subdivided into a training sample used to train the MVA and a test sample used to determine the separation power of the full analysis. The dashed line denotes the median of the photon test sample, which is used as the photon candidate cut. The inlay shows a zoom on the data distribution around the photon candidate cut.
}
\label{fig:bdt}
\end{figure}

In the next step, the MVA is applied to the $2{,}204$ events from the
data sample. The distribution of $\beta$ obtained for the data
sample is shown in Fig.~\ref{fig:bdt} and compared to the
distributions for the simulated samples. As expected from the
distributions of the individual observables, on average
smaller, i.e., less photon-like, values of $\beta$ for the data sample than for the proton
sample are found.

Finally, we use the distribution of $\beta$ for the data sample to identify photon candidate events. As in~\citet{Aab:2016agp}, we use a photon candidate cut fixed to the median of the photon (test sample) distribution in the energy range $E_\gamma {>}\unit[2{\times}10^{17}]{eV}$, which is shown in Fig.~\ref{fig:bdt} as a dashed line. Zero events from the data sample have $\beta$ value above the candidate cut value, hence no photon candidate events are identified. When looking at the events closest to the candidate cut, it can be noticed that their $S_b$ values are located towards the ``photon-like'' tail of the distribution for primary protons at the respective energies, typically at 1.5 to 2 standard deviations from the corresponding mean values for protons. Their $X_{\text{max}}$ values, however, are usually within one standard deviation from the corresponding average for protons. Regarding $N_{\text{stations}}$, a similar behaviour as for $S_b$ is found. In the combination of the individual observables in the MVA, the resulting value of $\beta$ is below the photon candidate cut.

\begin{figure}[b]
\plotone{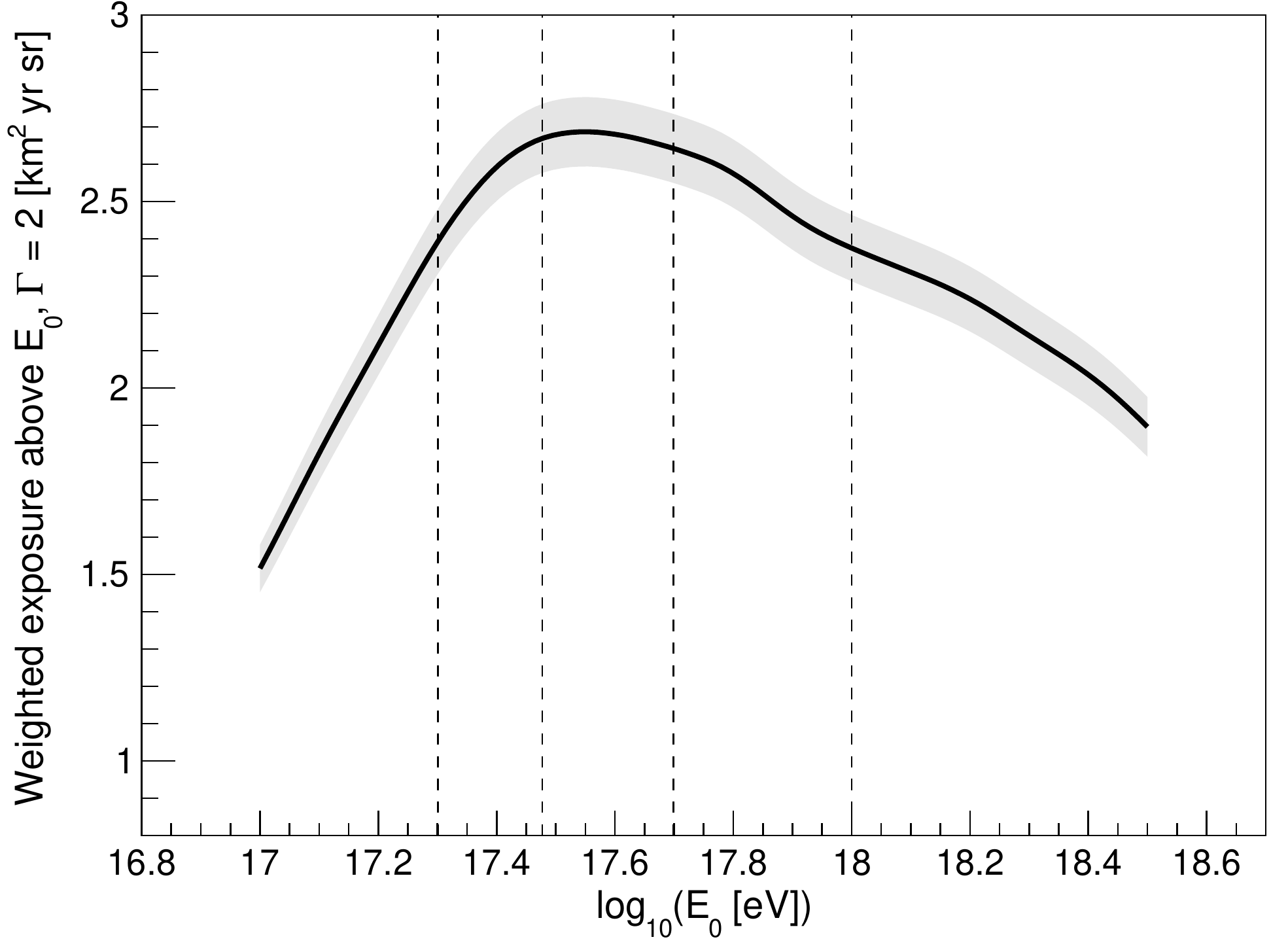}
\caption{
Integrated efficiency-weighted hybrid exposure for photons, calculated from simulations following Eq.~\ref{eq:exposure} under the assumption $\Gamma = 2$, with the statistical uncertainties shown as a grey band. The dashed lines denote the energy thresholds at which upper limits on the integral photon flux are placed in this analysis.
}
\label{fig:exposure}
\end{figure}

We calculate the final results of this study in terms of
upper limits on the integral flux of photons
$\Phi^{\text{C.L.}}_{\gamma,\,\text{U.L.}}(E_\gamma{>}E_0)$, where
C.L. denotes the confidence level at which we determine the upper
limits. $\Phi^{\text{C.L.}}_{\gamma,\,\text{U.L.}}(E_\gamma{>}E_0)$ is calculated according to 
\begin{equation}
\Phi^{\text{C.L.}}_{\gamma,\,\text{U.L.}}(E_\gamma{>}E_0) =
\frac{N^{\text{C.L.}}_\gamma}{\epsilon_{\text{cand}}{\times} (1-f_{\text{burnt}}){\times}\mathcal{E}_\gamma},
\label{eq:upperlimits}
\end{equation}
where $N^{\text{C.L.}}_\gamma$ is the upper limit on the number of
photon candidate events at the given confidence level calculated
using the Feldman-Cousins approach~\citep{Feldman:1997qc} with no
background subtraction, $\epsilon_{\text{cand}}$ is the efficiency of the photon candidate cut,
$f_{\text{burnt}}$ is the fraction of the data used as a ``burnt sample'', and $\mathcal{E}_\gamma$
is the integrated efficiency-weighted exposure for photons (see also Tab.~\ref{tab:results}). $\mathcal{E}_\gamma$ is calculated from
simulations as
\begin{equation}
\mathcal{E}_\gamma(E_\gamma{>}E_0) =
\int_{E_0}^{\infty} \frac{E_\gamma^{-\Gamma}}{c_E}\,
\epsilon_\gamma(E_\gamma, t, \theta, \varphi, x, y) \,\text{d}S \,\text{d}t \,
\text{d}\Omega \, \text{d}E_\gamma, 
\label{eq:exposure}
\end{equation}
where $\epsilon_\gamma(E_\gamma, t, \theta, \varphi, x, y)$ is the
overall efficiency for photons---excluding the final photon candidate cut---depending on the photon energy
$E_\gamma$, the time $t$, the zenith angle $\theta$, the azimuth angle $\varphi$ and
the coordinates $x$ and $y$ of the impact point of the air shower on
the ground. The integration is
performed over the area $S$, the time $t$,
the solid angle $\Omega$, and the photon energy $E_\gamma$. The normalization factor $c_E$ is calculated through
\begin{equation}
c_E = \int_{E_0}^{\infty} E_{\gamma}^{-\Gamma}\,\text{d}E_\gamma.
\end{equation}
The result of the integration following Eq.~\ref{eq:exposure}
with a spectral index $\Gamma = 2$ is shown in
Fig.~\ref{fig:exposure}. In the energy range of interest between $\unit[2{\times}10^{17}]{eV}$ and $\unit[10^{18}]{eV}$, the weighted exposure varies between $2.4$ and $\unit[2.7]{km^2\,yr\,sr}$, with a maximum at $\unit[3.5{\times}10^{17}]{eV}$. Towards lower energies, the exposure becomes smaller because lower-energy air showers trigger the detector with reduced efficiency. Towards higher energies, the dominant cause for the decrease in exposure is the event selection, because showers where $X_{\text{max}}$ is reconstructed to be below the field of view of the telescopes are excluded from the analysis (see Sec.~\ref{sec:samples}).

We place upper limits on the integral photon flux $\Phi^{\text{C.L.}}_{\gamma,\,\text{U.L.}}(E_\gamma{>}E_0)$ at threshold energies of $2$, $3$, and $\unit[5{\times}10^{17}]{eV}$, as well as $\unit[10^{18}]{eV}$, at a confidence level of $\unit[95]{\%}$. At these threshold energies, the upper limits are $2.72$, $2.50$, $2.74$, and $\unit[3.55]{km^{-2}\,yr^{-1}\,sr^{-1}}$, respectively. The quantities needed to calculate the upper limits according to Eq.~\ref{eq:upperlimits} are listed in Tab.~\ref{tab:results}. For completeness, we also calculated upper limits at a confidence level of $\unit[90]{\%}$, as used e.g. in \citet{Apel:2017ocm} and~\citet{Fomin:2017ypo}. The upper limits in this case are $2.15$, $1.97$, $2.16$, and $\unit[2.79]{km^{-2}\,yr^{-1}\,sr^{-1}}$ at the same threshold energies. Using the energy spectrum of cosmic rays measured by the Pierre Auger Observatory~\citep{PierreAuger:2021hun}, the upper limits on the integral photon flux can be translated into upper limits on the integral photon fraction. At a confidence level of $\unit[95]{\%}$, these are $\unit[0.28]{\%}$, $\unit[0.63]{\%}$, $\unit[2.20]{\%}$ and $\unit[13.8]{\%}$ for the same threshold energies as above.

\begin{table*}[t]
\caption{
Upper limits on the integral photon flux, determined at $\unit[95]{\%}$ C.L., calculated using Eq.~\ref{eq:upperlimits}. See the text for explanations.
}
\begin{tabular}{cccccccc}
\hline\hline
 $\unit[E_0]{[eV]}$ & $N_{\text{events}}$ & $N_\gamma$ & $N^{\unit[95]{\%}}_\gamma$ &
                            $\epsilon_{\text{cand}}$ & $1 - f_{\text{burnt}}$ &$\unit[\mathcal{E}_\gamma]{[km^2\,yr\,sr]}$
    &
      
      $\unit[\Phi^{\unit[95]{\%}}_{\gamma,\,\text{U.L.}}]{[km^{-2}\,yr^{-1}\,sr^{-1}]}$\\
    \hline
    $2{\times}10^{17}$ & 2,204 & 0 & 3.095 & 0.50 & 0.96 & 2.38 & 2.72\\
    $3{\times}10^{17}$ & 1,112 & 0 & 3.095 & 0.48 & 0.96 & 2.69 & 2.50\\
    $5{\times}10^{17}$ & 333 & 0 & 3.095 & 0.45 & 0.94 & 2.68 & 2.74\\
    $10^{18}$ & 67 & 0 & 3.095 & 0.38 & 0.94 & 2.41 & 3.55\\
\hline\hline    
\end{tabular}
\label{tab:results}
\end{table*}

To assess the impact on the final results of the choice of hadronic interaction models and of the assumptions on the composition of primary cosmic rays, smaller samples of proton-induced air showers simulated with the hadronic interaction models QGSJET-II-04~\citep{Ostapchenko:2010vb} and SIBYLL 2.3c~\citep{Fedynitch:2018cbl} and of air showers induced by iron nuclei, simulated with EPOS LHC, have been used. Each of these samples contains $30{,}000$ air-shower events. The analysis has been repeated replacing the default background sample (primary protons simulated with EPOS LHC) by primary protons simulated with QGSJET-II-04 and SIBYLL 2.3c and with a mixture of $\unit[50]{\%}$ primary protons and $\unit[50]{\%}$ primary iron nuclei (both simulated with EPOS LHC). In all cases, no photon candidate events were identified in the data sample, indicating that the analysis is robust against these assumptions. Likewise, varying the spectral index $\Gamma$ from 2 to, e.g., $1.5$ or $2.5$, and repeating the analysis does not change the observed number of photon candidates (0). It should be taken into account however, that $\Gamma$ also enters the calculation of the weighted exposure, leading to a change in the final upper limits by, on average, $\unit[5]{\%}$. Finally, we studied the impact of possible systematic uncertainties in the measurement of the observables. Changing the $X_{\text{max}}$ values of all events in the data sample by $\unit[\pm 10]{g\,cm^{-2}}$~\citep{bellido:2017lem} does not change the number of photon candidate events. Likewise, changing the $S_b$ values of all events in the data sample by $\unit[\pm 5]{\%}$~\citep{Aab:2016agp} has no effect on the number of photon candidate events. These tests show that the analysis is also robust against systematic uncertainties in the measured observables.

\section{Discussion and Conclusions}
\label{sec:discussion}

\begin{figure*}[t]
\epsscale{0.9}
\plotone{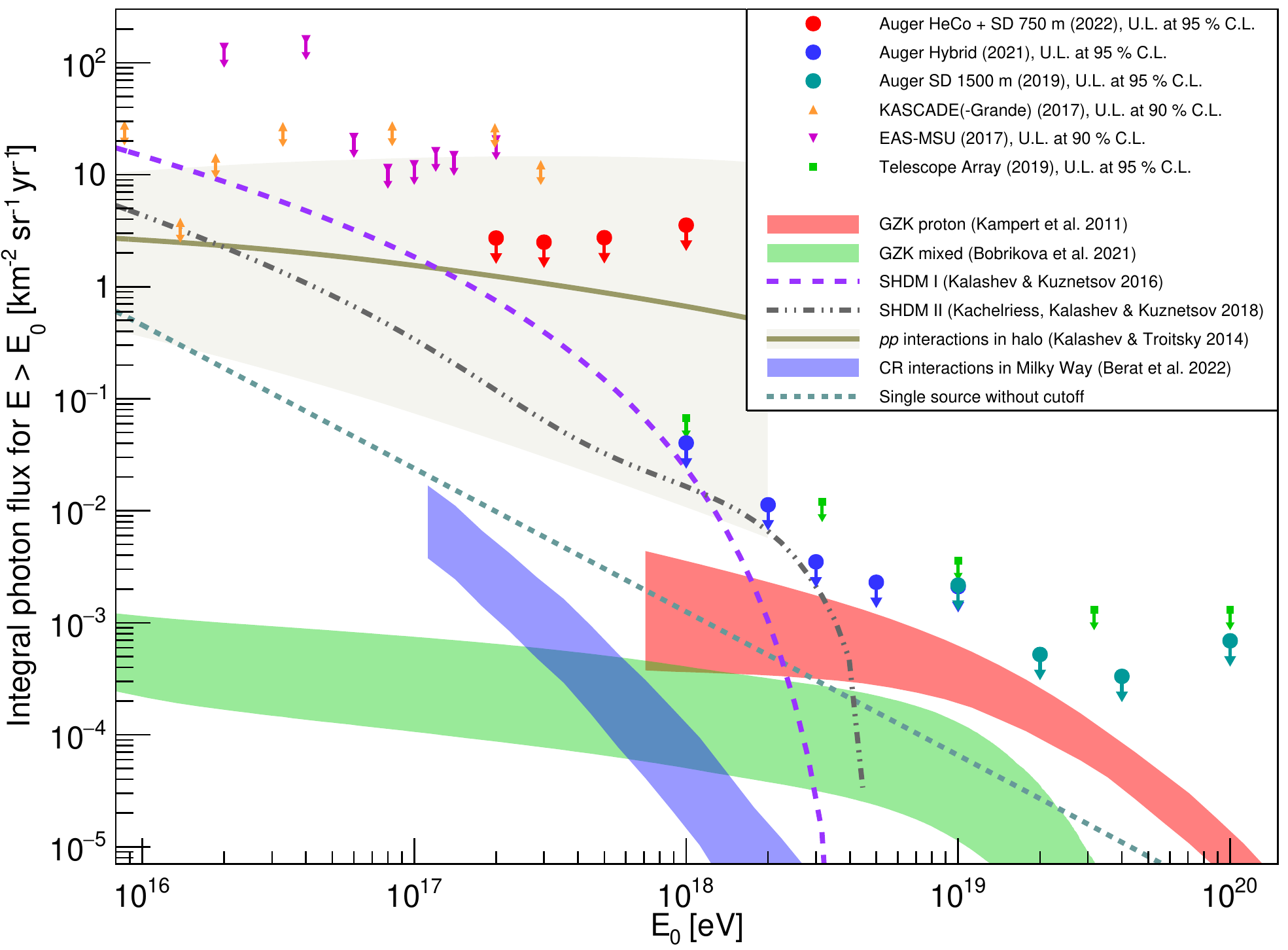}
\caption{
Upper limits (at $\unit[95]{\%}$ C.L.) on the integral photon flux above $\unit[2{\times}10^{17}]{eV}$ determined here (red circles). Shown are also previous upper limits by various experiments: Pierre Auger Observatory (hybrid: blue circles, taken from~\citet{Savina:2021zrn}; SD: cyan circles, taken from~\citet{Rautenberg:2019TI}), KASCADE/KASCADE-Grande (orange triangles, taken from~\citet{Apel:2017ocm}), EAS-MSU (magenta diamonds, taken from~\citet{Fomin:2017ypo}) and Telescope Array (green squares, taken from~\citet{Abbasi:2018ywn}). The red band denotes the range of expected GZK photon fluxes under the assumption of a pure-proton scenario~\citep{Kampert:2011hkm}. The green band shows the expected GZK photon flux assuming a mixed composition that would fit the Auger data~\citep{Bobrikova:2021kuj}. In addition, the expected photon fluxes from the decay of super-heavy dark matter particles are included (decay into hadrons: dashed violet line, based on~\citet{Kalashev:2016cre}; decay into leptons: dot-dashed gray line, based on~\citet{Kachelriess:2018rty}; the exact lines have been obtained through personal communication with one of the authors). The photon fluxes that would be expected from $pp$ interactions in the Galactic halo~(\citet{Kalashev:2014vra}, olive-green line) or from cosmic-ray interactions with matter in the Milky Way~(\citet{Berat2021}, blue band) are shown as well. Also included is the expected flux of photons from a single, putative source without a cutoff in its spectrum (dotted turquoise line, modeled after HAWC J1825-134,~\citet{Albert:2020yty}, where we extrapolated the measured flux to the highest energies), ignoring its directionality as if its flux were distributed over the full sky.
}
\label{fig:UL}
\end{figure*}

The upper limits on the integral photon flux derived in the previous section are shown in Fig.~\ref{fig:UL}, together with the results of other photon searches with energy thresholds ranging from $\unit[10^{16}]{eV}$ to $\unit[10^{20}]{eV}$. In the energy region below $\unit[10^{18}]{eV}$, the limits obtained in this study are the most stringent ones, improving previous limits from KASCADE-Grande~\citep{Apel:2017ocm} and EAS-MSU~\citep{Fomin:2017ypo} by up to an order of magnitude. The analysis presented here extends the energy range of photon searches at the Pierre Auger Observatory and complements previous analyses in the energy range above $\unit[10^{19}]{eV}$~\citep{Rautenberg:2019TI} and between $\unit[10^{18}]{eV}$ and $\unit[10^{19}]{eV}$~\citep{Savina:2021zrn}, closing the gap to the smaller air-shower experiments mentioned before. For a threshold energy of $\unit[10^{18}]{eV}$, the upper limit determined in this analysis is about two orders of magnitude above the previous limit from ~\citet{Savina:2021zrn}, which is due mainly to the smaller exposure, which in turn is a consequence of the smaller size of the $\unit[750]{m}$ SD array compared to the full array. Overall, the Pierre Auger Observatory now provides the most stringent limits on the incoming UHE photon flux over three decades in energy. This set of upper limits allows us to draw some conclusions relevant to the astrophysics of UHECRs and beyond, which we now discuss.

A guaranteed flux of ultra-high-energy photons of cosmogenic origin is that resulting from interactions of UHECRs with the background photon fields permeating the Universe, most notably the cosmic microwave background~\citep{Greisen:1966jv,Zatsepin:1966jv}. This flux is much reduced relative to that of UHECRs due to, as pointed out in the introduction, the short photon horizon (a few hundred kpc) compared to the cosmic-ray one (from a few tens of Mpc above the GZK threshold to cosmological scales below). Although guaranteed, the precise knowledge of this flux suffers from several uncertainties. The production channel of these photons is the decay of $\pi^0$ mesons. The hadrons that cause the creation of these mesons may be primary proton cosmic rays, or secondary ones mainly produced by the photo-disintegration of nuclei interacting inelastically with a cosmic background photon. Since the nucleons produced in a photo-disintegration reaction inherit the energy of the fragmented nucleus divided by its atomic number, the photons ultimately produced from primary heavy nuclei are of lower energies than those from lighter ones or from proton primaries. The photon flux, therefore, depends on the nature of the UHECRs, which remains poorly constrained above about $\unit[5{\times}10^{19}]{eV}$. The expectation for a pure-proton scenario is shown as the red band in Fig.~\ref{fig:UL}~\citep{Kampert:2011hkm}, while that for a mixed composition at the sources is shown as the green band~\citep{Bobrikova:2021kuj}. The latter, which is an order of magnitude lower than the former and falls off much faster, is in agreement with the various constraints inferred from the data collected at the Observatory, namely the mass composition and the energy spectrum~\citep{PierreAuger:2016use}. Other dependencies that explain the width of the bands come from the hypotheses on the maximum acceleration energy of the nuclei at the sources and on the shape of the energy spectrum of the accelerated particles. Overall, while the sensitivities reached above about $\unit[3{\times}10^{18}]{eV}$ approach the most optimistic expectations of the cosmogenic photon flux from protons, they are about $1.5$ orders of magnitude above those from the mixed-composition model. 

Another cosmogenic flux is that from the interactions of UHECRs with the matter traversed in the Galactic plane, which is larger than the aforementioned one below about $\unit[10^{18}]{eV}$~\citep{Berat2021}. Shown in blue, the width of the band accounts for uncertainties arising from the distribution of the gas in the disk, the absolute level of the UHECR  flux, and the mass composition. The limits obtained in this study improve previous ones in the energy range of interest to probe such a flux; yet they remain between two and three orders of magnitude above the expectations. 

The cosmogenic fluxes just mentioned can be seen as floors above which increased sensitivity to photons could reveal unexpected phenomena. To exemplify such a potential, we explain below the four curves that correspond to fluxes from putative sources in the Galactic disk or to patterns that could emerge from proton-proton interactions in the halo of the Galaxy or from the decay of super-heavy dark matter (SHDM). 

The recent observation of photons above $\unit[2{\times}10^{14}]{eV}$ from decaying neutral pions from the J1825-134 source reported in~\citet{Albert:2020yty}, in a direction coincident with a giant molecular cloud, provides evidence that cosmic rays are indeed accelerated to energies of several $\unit[10^{15}]{eV}$, and above, in the Galaxy. Interestingly, the flux of this source could extend well beyond $\unit[2{\times}10^{14}]{eV}$, as no cutoff is currently observed in its energy spectrum measured up to this energy. As an example of the discovery potential with increased exposure, we show as the green curve the flux from such a putative source extrapolated to the highest energies. Note that this flux, which is directional in essence, is here for simplicity calculated by converting it to a diffuse one, assuming the flux were distributed over the full sky. We observe that the extrapolated flux for this source is higher than the cosmogenic ones below $\unit[10^{18}]{eV}$. The upper limits determined here exceed the extrapolated flux of this single, specific source by two orders of magnitude. They nevertheless limit the effective number of similar sources in the Galaxy. Improved tests of the abundance of such putative sources will be possible by further increasing the sensitivity of photon searches in this energy region or decreasing the energy threshold.

The origin of the bulk of the high-energy neutrino flux observed at the IceCube observatory (see, e.g., \citet{IceCube:2020acn}) is still debated. However, their production mechanism is conventionally considered as that of high-energy hadronic or photo-hadronic interactions that create charged pions decaying into neutrinos. These same interactions produce neutral pions that decay into photons. Therefore, there is an expected connection between high-energy photons and high-energy neutrinos. Since the horizon of photons is much smaller than that of neutrinos, they can trace the local sources in a way that could facilitate the differentiation between different scenarios. In Fig.~\ref{fig:UL}, we reproduce in olive green the expectations for cosmic-ray interactions with the hot gas filling the outer halo of the Galaxy up to hundreds of kiloparsecs, as estimated in~\citet{Kalashev:2014vra} by requiring that this photon flux is the counterpart of the neutrino one. The width of the band reflects the uncertainties in the spectral shape of the neutrino flux. We observe that the limits derived in this study are already constraining.

Finally, UHE photons could also result from the decay of SHDM particles. We note that previous upper limits on the incoming photon flux already severely constrained non-acceleration models in general, and SHDM models in particular, trying to explain the origin of cosmic rays at the highest energies (see, e.g.,~\citet{PierreAuger:2007hjd,Aab:2016agp}). Still, the production of super-heavy particles in the early Universe remains a possible solution to the dark matter conundrum because of the high value of the instability energy scale in the Standard Model of particle physics, which, according to current measurements of the Higgs-boson mass and the Yukawa coupling of the top quark, ranges between $10^{10}$ and $\unit[10^{12}]{GeV}$~\citep{Degrassi:2012ry,Bednyakov:2015sca}. The Standard Model can, therefore, be extrapolated without encountering inconsistencies that would make the electroweak vacuum unstable up to such energy scales (and even to much higher ones given the slow evolution of the instability scale up to the Planck mass~\citep{Degrassi:2012ry}), where new physics could arise, giving rise to a mass spectrum of super-heavy particles that could have been produced during post-inflation reheating by various mechanisms (see, e.g.,~\citet{Ellis:1990nb,PhysRevLett.79.4302,Chung:1998zb,Garny:2015sjg,Ellis:2015jpg,Dudas:2017rpa,Kaneta:2019zgw,Mambrini:2021zpp}). The set of limits shown in Fig.~\ref{fig:UL} allows for constraining the phase space of mass and lifetime of the SDHM particles (see, e.g.,~\citet{Kalashev:2016cre,Kachelriess:2018rty,Berat2021}). To illustrate the  discovery potential with searches for UHE photons, we show as the dashed violet line and the dot-dashed gray line the expected photon fluxes in the case of hadronic~\citep{Kalashev:2016cre} and leptonic~\citep{Kachelriess:2018rty} decay channels, respectively. For these lines, we assume that the mass $M_X$ of the SHDM particles is $\unit[10^{10}]{GeV}$ and their lifetime $\tau_X$ is $\unit[3{\times}10^{21}]{yr}$, as currently allowed by previous limits. As the sensitivity of current photon searches increases, it will be possible to further constrain these values~\citep{Anchordoqui:2021crl}.

Further improvements of the upper limits derived in this analysis can be expected not only from using a larger dataset, profiting from the constant increase in exposure over time, but also from the ongoing detector upgrade of the Pierre Auger Observatory, dubbed AugerPrime~\citep{Castellina:2019irv,PierreAuger:2016qzd}. A major part of this upgrade is the installation of scintillation detectors on top of the water-Cherenkov detector stations of the SD, with the aim to better separate the muonic and electromagnetic components of an air shower. Current photon searches already exploit the differences in these components between photon- and hadron-induced air showers, albeit in a rather indirect way. AugerPrime will allow for a more direct access, which will lead to an overall better separation between photon-induced air showers and the vast hadronic background. Naturally, this upgrade will improve the upper limits on the incoming photon flux or, in the best case, lead to the unambiguous detection of photons at ultra-high energies.

\section*{Acknowledgments}

\begin{sloppypar}
The successful installation, commissioning, and operation of the Pierre
Auger Observatory would not have been possible without the strong
commitment and effort from the technical and administrative staff in
Malarg\"ue. We are very grateful to the following agencies and
organizations for financial support:
\end{sloppypar}

\begin{sloppypar}
Argentina -- Comisi\'on Nacional de Energ\'\i{}a At\'omica; Agencia Nacional de
Promoci\'on Cient\'\i{}fica y Tecnol\'ogica (ANPCyT); Consejo Nacional de
Investigaciones Cient\'\i{}ficas y T\'ecnicas (CONICET); Gobierno de la
Provincia de Mendoza; Municipalidad de Malarg\"ue; NDM Holdings and Valle
Las Le\~nas; in gratitude for their continuing cooperation over land
access; Australia -- the Australian Research Council; Belgium -- Fonds
de la Recherche Scientifique (FNRS); Research Foundation Flanders (FWO);
Brazil -- Conselho Nacional de Desenvolvimento Cient\'\i{}fico e Tecnol\'ogico
(CNPq); Financiadora de Estudos e Projetos (FINEP); Funda\c{c}\~ao de Amparo \`a
Pesquisa do Estado de Rio de Janeiro (FAPERJ); S\~ao Paulo Research
Foundation (FAPESP) Grants No.~2019/10151-2, No.~2010/07359-6 and
No.~1999/05404-3; Minist\'erio da Ci\^encia, Tecnologia, Inova\c{c}\~oes e
Comunica\c{c}\~oes (MCTIC); Czech Republic -- Grant No.~MSMT CR LTT18004,
LM2015038, LM2018102, CZ.02.1.01/0.0/0.0/16{\textunderscore}013/0001402,
CZ.02.1.01/0.0/0.0/18{\textunderscore}046/0016010 and
CZ.02.1.01/0.0/0.0/17{\textunderscore}049/0008422; France -- Centre de Calcul
IN2P3/CNRS; Centre National de la Recherche Scientifique (CNRS); Conseil
R\'egional Ile-de-France; D\'epartement Physique Nucl\'eaire et Corpusculaire
(PNC-IN2P3/CNRS); D\'epartement Sciences de l'Univers (SDU-INSU/CNRS);
Institut Lagrange de Paris (ILP) Grant No.~LABEX ANR-10-LABX-63 within
the Investissements d'Avenir Programme Grant No.~ANR-11-IDEX-0004-02;
Germany -- Bundesministerium f\"ur Bildung und Forschung (BMBF); Deutsche
Forschungsgemeinschaft (DFG); Finanzministerium Baden-W\"urttemberg;
Helmholtz Alliance for Astroparticle Physics (HAP);
Helmholtz-Gemeinschaft Deutscher Forschungszentren (HGF); Ministerium
f\"ur Innovation, Wissenschaft und Forschung des Landes
Nordrhein-Westfalen; Ministerium f\"ur Wissenschaft, Forschung und Kunst
des Landes Baden-W\"urttemberg; Italy -- Istituto Nazionale di Fisica
Nucleare (INFN); Istituto Nazionale di Astrofisica (INAF); Ministero
dell'Istruzione, dell'Universit\'a e della Ricerca (MIUR); CETEMPS Center
of Excellence; Ministero degli Affari Esteri (MAE); M\'exico -- Consejo
Nacional de Ciencia y Tecnolog\'\i{}a (CONACYT) No.~167733; Universidad
Nacional Aut\'onoma de M\'exico (UNAM); PAPIIT DGAPA-UNAM; The Netherlands
-- Ministry of Education, Culture and Science; Netherlands Organisation
for Scientific Research (NWO); Dutch national e-infrastructure with the
support of SURF Cooperative; Poland -- Ministry of Education and
Science, grant No.~DIR/WK/2018/11; National Science Centre, Grants
No.~2016/22/M/ST9/00198, 2016/23/B/ST9/01635, and 2020/39/B/ST9/01398;
Portugal -- Portuguese national funds and FEDER funds within Programa
Operacional Factores de Competitividade through Funda\c{c}\~ao para a Ci\^encia
e a Tecnologia (COMPETE); Romania -- Ministry of Research, Innovation
and Digitization, CNCS/CCCDI -- UEFISCDI, projects PN19150201/16N/2019,
PN1906010, TE128 and PED289, within PNCDI III; Slovenia -- Slovenian
Research Agency, grants P1-0031, P1-0385, I0-0033, N1-0111; Spain --
Ministerio de Econom\'\i{}a, Industria y Competitividad (FPA2017-85114-P and
PID2019-104676GB-C32), Xunta de Galicia (ED431C 2017/07), Junta de
Andaluc\'\i{}a (SOMM17/6104/UGR, P18-FR-4314) Feder Funds, RENATA Red
Nacional Tem\'atica de Astropart\'\i{}culas (FPA2015-68783-REDT) and Mar\'\i{}a de
Maeztu Unit of Excellence (MDM-2016-0692); USA -- Department of Energy,
Contracts No.~DE-AC02-07CH11359, No.~DE-FR02-04ER41300,
No.~DE-FG02-99ER41107 and No.~DE-SC0011689; National Science Foundation,
Grant No.~0450696; The Grainger Foundation; Marie Curie-IRSES/EPLANET;
European Particle Physics Latin American Network; and UNESCO.
\end{sloppypar}

\bibliographystyle{aasjournal}
\bibliography{references}

\end{document}